%% file: ss433sub2.tex
\documentclass[twocolumn,tighten,times]{aastex631}

\usepackage{color}
\usepackage{url}
\usepackage{hyperref}
\usepackage{xspace}
\usepackage{soul}

\shorttitle{X-ray polarization of SS~433}
\shortauthors{Kaaret et al.}

\begin{document}

\title{X-ray Polarization of the Eastern Lobe of SS~433}
\include{author_list}


\begin{abstract}
How astrophysical systems translate the kinetic energy of bulk motion into the acceleration of particles to very high energies is a pressing question. SS~433 is a microquasar that emits TeV $\gamma$-rays indicating the presence of high-energy particles. A region of hard X-ray emission in the eastern lobe of SS~433 was recently identified as an acceleration site. We observed this region with the Imaging X-ray Polarimetry Explorer and measured a polarization degree in the range 38\% to 77\%. The high polarization degree indicates the magnetic field has a well ordered component if the X-rays are due to synchrotron emission. The polarization angle is in the range $-12\arcdeg$ to $+10\arcdeg$ (east of north) which indicates that the magnetic field is parallel to the jet. Magnetic fields parallel to the bulk flow have also been found in supernova remnants and the jets of powerful radio galaxies. This may be caused by interaction of the flow with the ambient medium.
\end{abstract}



\section{Introduction}
\label{sec:intro}

A variety of astrophysical systems are capable of translating the kinetic energy of bulk motion into the acceleration of particles to very high energies. Some examples include supernova remnants \citep{Vink2012}, the jets of supermassive black holes and stellar-mass compact objects \citep{Blandford2019,Mirabel1999}, and explosive events such as $\gamma$-ray bursts \citep{Kumar2015}. Understanding the physical conditions within the acceleration regions is key to deciphering the acceleration process.

SS~433 is a stellar binary system exhibiting jets that generate TeV $\gamma$-rays revealing the presence of energetic particles \citep{Margon1984,HAWC2018}. SS~433 contains a stellar-mass compact object (likely a black hole) accreting from a supergiant companion \citep{Fabrika2004} and lies within the Milky Way at a distance of 5.5~kpc \citep{Hjellming1981,Blundell2004}. The compact object launches bipolar jets (with a position angle of $100\arcdeg$ and inclined $80\arcdeg$ to our line of sight) containing ionized matter moving at a bulk speed of $0.26c$ nearly perpendicular to our line of sight \citep{Margon1984}. The jets precess with a period of 162.5~days and an opening angle of $20\arcdeg$ and interact with the surrounding supernova remnant, W50, creating two lobes east and west from SS~433 and a unique morphology denoted as the `Manatee'.\footnote{\url{https://www.nrao.edu/pr/2013/w50/}} 

The jet X-ray emission has a thermal spectrum with prominent line emission within 0.16~pc of SS~433 \citep{Marshall2002,Migliari2002}. The spectrum is non-thermal in the east (``e1") and west (``w1") lobes starting about 30~pc from SS~433 \citep{SafiHarb1997,Yamauchi1994,SafiHarb1999}. Soft thermal emission is seen from the termination shock at the jet end \citep{SafiHarb1997,Brinkmann2007}. The X-ray spectra of the e1 and w1 lobes are consistent with synchrotron radiation from energetic electrons \citep{Watson1983,SafiHarb1999}. The High Altitude Water Cherenkov (HAWC) observatory, in Mexico, detected very high-energy (VHE) $\gamma$-ray emission from the e1 and w1 lobes extending up to at least 25~TeV indicating the presence of particles with energies of 100's of TeV \citep{HAWC2018}. The VHE gamma-ray emission has been confirmed and more accurately localized with the Large High Altitude Air Shower Observatory (LHAASO) \citep{LHASSO2023}. Detection of gamma-ray emission from the lobes with the High Energy Stereoscopic System (H.E.S.S.) has also been reported \citep{OliveraNieto2022}. The TeV emission and the non-thermal X-ray emission is localized to the lobes which suggests that particle acceleration occurs in the jet at significant distances away from SS~433.

\begin{figure*}[tb] 
\centerline{\includegraphics[width=0.8\textwidth]{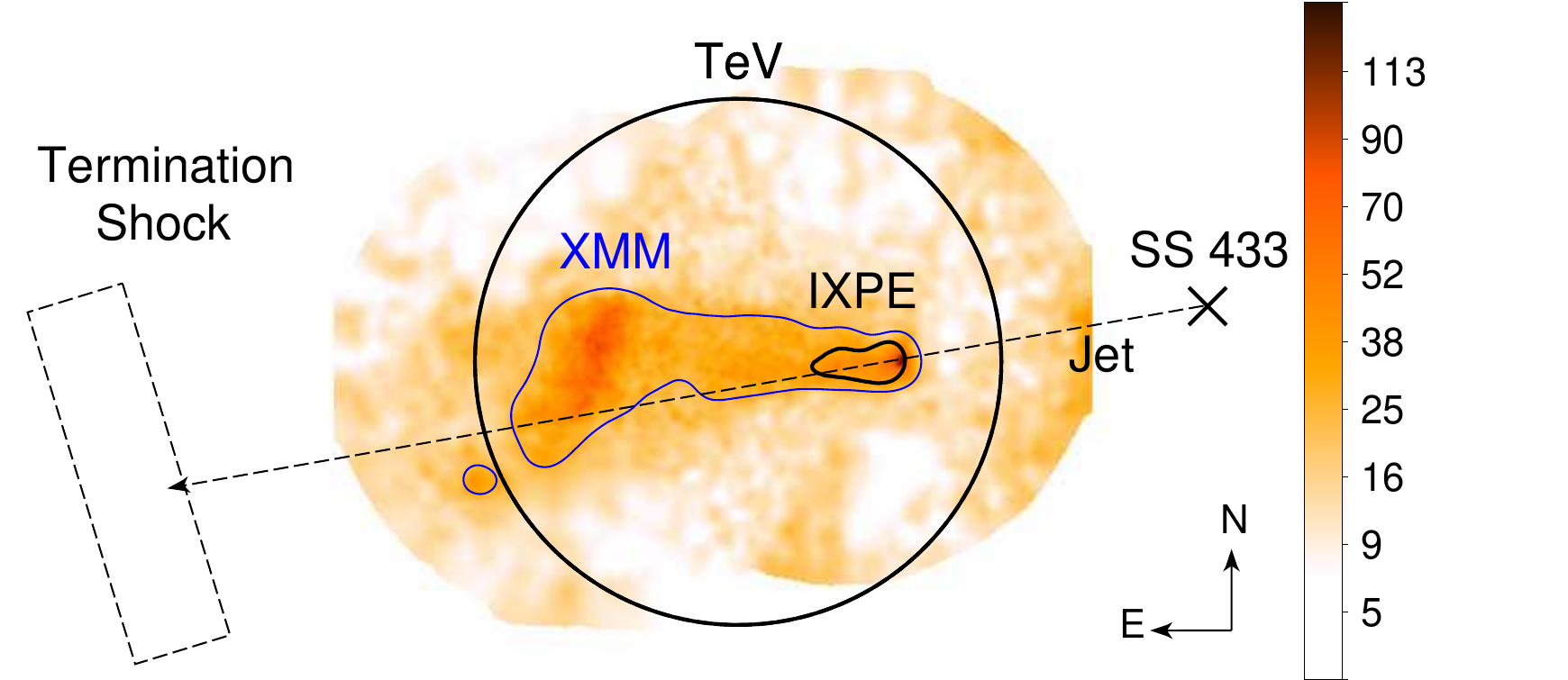}} \vspace{12pt}
\centerline{\includegraphics[width=0.8\textwidth]{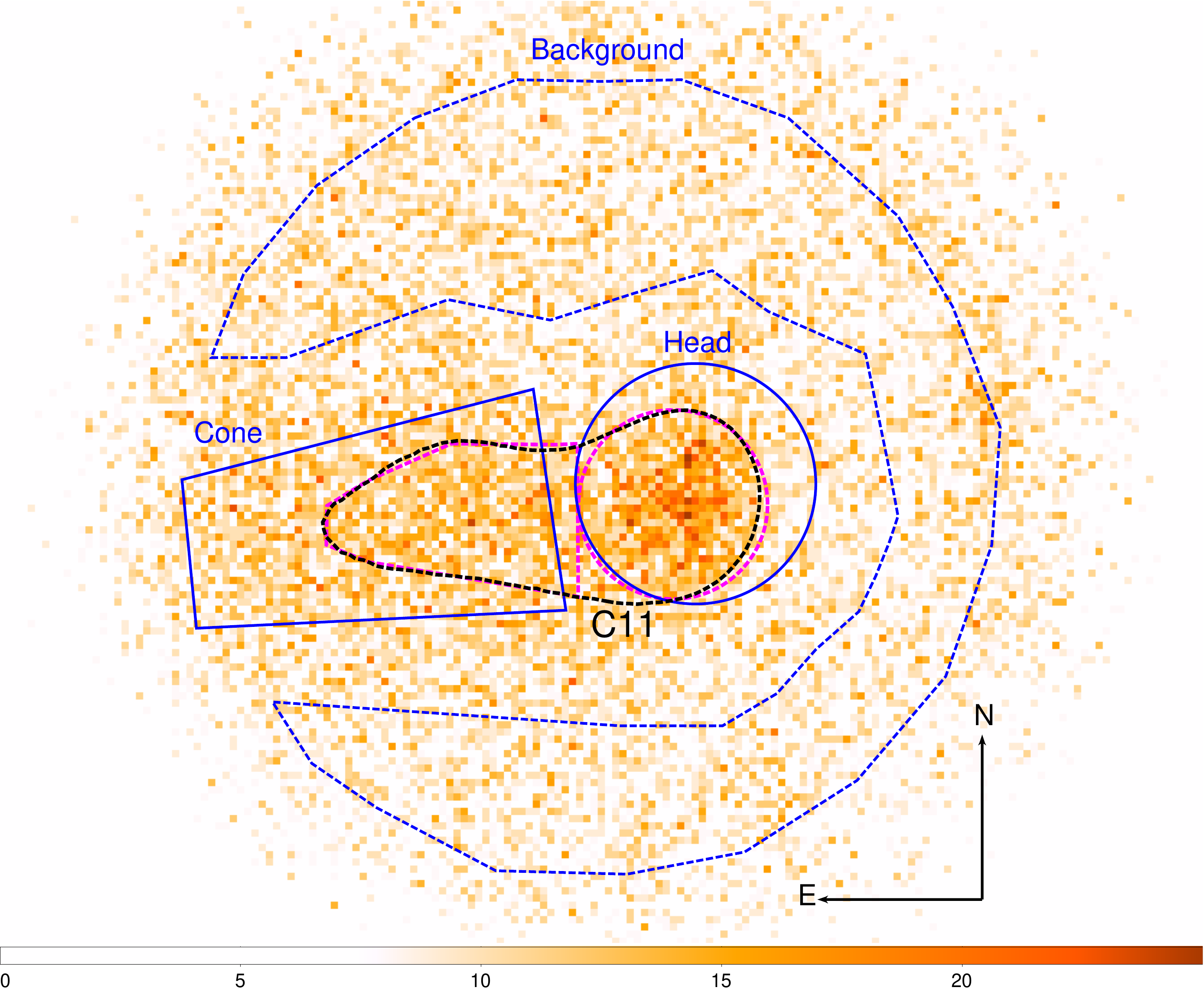}}
\caption{X-ray images of the eastern lobe of SS~433. (Top) Mosaic of {XMM-Newton} data in the 0.3--10~keV band. The $\times$ marks the location of the binary. The dashed line indicates the jet axis \citep{Hjellming1981} and the dashed rectangle marks the jet termination shock. The blue contour is 25 counts~pixel$^{-1}$ in the {XMM-Newton} image. The black contour is the {IXPE} source region. The black circle is the HAWC error circle and has a radius of $15\arcmin$. (Bottom) X-ray image from {IXPE} in the 2--5~keV band. C11 marks the black contour used as the {IXPE} source region. The partially-overlapping magneta contours indicate the forward region (overlapping the Head) and trailing region (overlapping the Cone) used for spatially-resolved polarization analysis. The blue, dashed contour marks the background region.  The blue circle is the Head region and the blue quadrilateral is the Cone region from \citet{SafiHarb2022}. The black arrows have a length of~2\arcmin (3.2~pc).} \label{fig:xray_images}
\end{figure*}

Using XMM-Newton and NuSTAR, \citet{SafiHarb2022} identified a region (the `Head', see the lower panel of Fig.~\ref{fig:xray_images}) of nonthermal hard X-ray emission (3--30~keV) at $\sim 18\arcmin$ (29~pc) east of SS~433 coincident with the onset of X-ray emission in the e1 lobe closest to SS~433. Assuming the hard X-rays are synchrotron radiation, the hard spectrum (photon index $\Gamma \approx 1.6$) suggests that this is the site of the acceleration of the high energy particles. This is supported by the morphology of the X-ray emission which shows the hardest X-ray emission closest to SS~433, with lower energy X-rays forming a cometary tail pointed away from SS~433. The 'Head' does not appear to have counterparts at longer wavelengths, thus no polarization information is available.

We observed the Head region using the {Imaging X-ray Polarimetry Explorer} ({IXPE}) in order to measure the polarization of the X-ray emission and constrain the geometry of the magnetic field. We describe the observations and our analysis in Section~\ref{sec:data} and present the X-ray polarization results in Section~\ref{sec:results}. We discuss the results and the implications for the acceleration mechanism in Section~\ref{sec:discussion}.

\section{Observations and Data Reduction}
\label{sec:data}

{IXPE} is the first satellite mission dedicated to X-ray polarimetry in the 2--8 keV band \citep{Weisskopf2022}. It contains three X-ray telescopes, each consisting of a Mirror Module Assembly \citep{Ramsey2022} and a polarization-sensitive gas-pixel detector unit (DU) \citep{Costa2001,Soffitta2021,Baldini2021}, that enable imaging X-ray polarimetry of extended sources. {IXPE} provides an angular resolution of $\sim 30\arcsec$ (half-power diameter). The overlap of the fields of view of the three DUs is circular with a diameter of $12\farcm{9}$. The DUs record the tracks of photoelectrons produced by X-rays to enable measurement of linear X-ray polarization.

{IXPE} observed the eastern lobe of SS~433 from 2023 April 19 to 2023 May 9 with a gap on 2023 May 2 due to observations of another target. The target coordinates (RA = $288\fdg27799$, DEC = $4\fdg93573$) were selected to cover the region of hard X-ray emission identified by \citet{SafiHarb2022}. The target is $5\farcm7$ from the LHASSO position and within the error circle.

Because our target is extended and relatively faint, the photon event list was processed to reduce the instrumental background that is induced by Earth albedo and primary energetic particles and secondaries produced in their interactions in the detectors or elsewhere on the spacecraft. The rejection algorithm is described in \citet{DiMarco2023} and removes $\sim$40\% of the background events and a negligible fraction ($\sim$1\%) of good events. The main components of the residual background are secondary positrons and a few primary protons that cannot be morphologically distinguished from X-ray initiated photoelectron tracks \citep{Xie2021}. We also removed times of high background found following bright solar flares (likely associated with coronal mass ejections) and surrounding intervals when {IXPE} crosses the South Atlantic Anomaly. These backgrounds appear as short term increases in counting rate and we removed times when the count rate was higher than the mean rate plus three times the rms of the rate \citep{Forsblom2023}. The technique has been important for the study of faint sources such as Galactic Center clouds and supernova remnants with IXPE \citep{Marin2023,Zhou2023}. After filtering, the total livetime was 800.7, 791.8, and 789.2~ks for DU1, DU2, and DU3, respectively. The background is higher at the field of view edges of each DU, but relatively uniform across the central $10\arcmin$ diameter \citep{DiMarco2023}. We limit our analysis to this central region.

We selected an energy band to optimize the signal from the source versus the background, both instrumental and from the cosmic X-ray background. We extracted a source spectrum using the `Head' region defined by \citet{SafiHarb2022} and a background spectrum from a nearby region without strong diffuse emission. Scaling for the region sizes, the source spectrum is above the background spectrum in the energy band of 2 to 5~keV, which we use for all subsequent analysis. 

The {IXPE} image in this band is shown in Fig.~\ref{fig:xray_images} along with an {XMM-Newton} mosaic in the 0.3--10~keV band \citep{SafiHarb2022}. We define an {IXPE} source region, that we refer to as `C11',  by drawing a contour corresponding to 11 counts per pixel in the {IXPE} map using the SAOImage DS9 image display tool with the smoothing parameter set to 16 \citep{ds9}. The region covers most of the `Head' and `Cone' regions. We define a background region (shown as a dashed line in Fig~\ref{fig:xray_images}) that avoids areas of strong diffuse emission and is within the central $10\arcmin$ diameter of the image. 

We extracted events within the source and background regions using \textsc{ixpeobssim} version 30.5.0 \citep{baldnini22}. We calculated polarization data cubes in the 2--5~keV energy band using the \texttt{pcube} algorithm in the \texttt{xpbin} tool and spectra of  $I$, $Q$, and $U$ Stokes parameters with PHA1, PHAQ1, and PHAU1 algorithms in \texttt{xpbin}. We performed spectropolarimetric analysis using \textsc{xspec} version 12.13.1. We used version 12 of {IXPE} instrument response files and performed an unweighted analysis for both \texttt{pcube} and \textsc{xspec}.

\begin{figure} 
\includegraphics[width=1.1\linewidth]{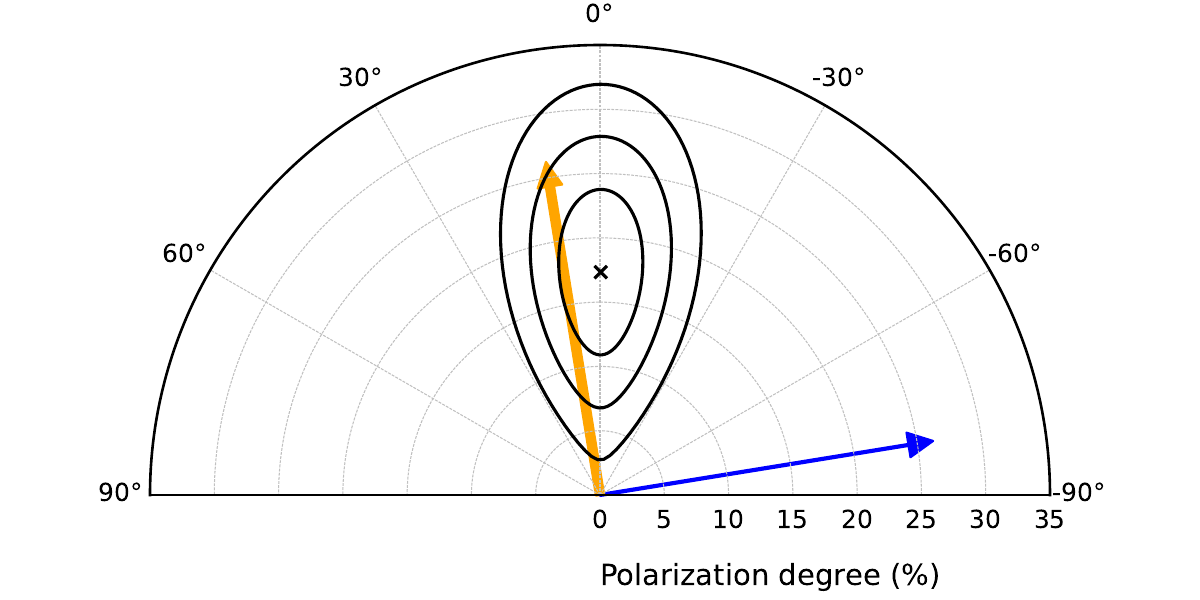}
\caption{Contour plots of the polarization degree and angle measured with \texttt{pcube}. Position angles are measured positive east of north. The $\times$ marks the measured values. The contours are at the 68.27\%, 95.45\%, and 99.73\% confidence intervals. The blue arrow indicates the direction from the lobe to the compact object while the orange arrow is perpendicular.} \label{fig:pcube_polar}
\end{figure}

\section{Polarization Results}
\label{sec:results}

Figure~\ref{fig:pcube_polar} shows the polarization of the events from the C11 region as measured using the model-independent \texttt{pcube} algorithm. The observed polarization degree (PD) is $17.3\% \pm 4.3\%$ and the electric vector position angle (EVPA) is $-0\fdg2 \pm  7\fdg0$ (all uncertainties are 68\% confidence). Position angles are measured anticlockwise from north in the equatorial coordinate system as per the convention of the International Astronomical Union. The probability that the signal is a random fluctuation from an unpolarized source is $2.4 \times 10^{-4}$. This is equivalent to a $3.7 \sigma$ confidence level. We note that the energy band and source region selection were based on the intensity data alone, thus there is only one trial for calculation of the significance of the polarization detection. The events include the X-ray emission from the lobe of SS~433, the cosmic X-ray background, and the residual instrumental background. All are expected to be unpolarized except for the emission from the lobe. Thus, the result indicates a highly probable detection of polarization in the X-ray emission of the lobe. 

Scaling from the background region, we estimate that 30\% of the counts are from the lobe and 70\% are from instrumental or astrophysical backgrounds. Utilizing the additive nature of the Stokes parameters, we calculated the Stokes $I$, $Q$, and $U$ for the source and background regions then subtracted the background from the source after scaling for the different region areas. After background subtraction, we estimate the PD of the lobe emission as $62\% \pm 15\%$ and the EVPA as $2\fdg8 \pm  7\fdg0$. As expected for unpolarized backgrounds, the EVPA is consistent with the results before background subtraction. The PD is consistent with that calculated using the polarization dilution equation (eq.\ 2) in  \citet{DiMarco2023}.

We also divided the C11 region into two parts, see Fig.~\ref{fig:xray_images}, and performed a background-subtracted polarization analysis. The circular region covering the forward region has $\rm PD = 59\% \pm 17\%$, $\rm EVPA = 6\arcdeg \pm 8\arcdeg$, and a chance probability of 0.0044. While the polygon covering the trailing region does not provide a significant detection, yielding a chance probability of 0.033, the PD ($48\% \pm 18\%)$ and EVPA ($4\arcdeg \pm 10\arcdeg$) are consistent with those of C11. The lower detection significance of the forward region relative to C11 suggests that the polarization signal is not concentrated in the forward region.

We measured the polarization of the events from the background region and found that it is consistent with zero. The same background region is used for the spectropolarimetic analysis described below. We also investigated three rectangular background regions placed to the North, West, and South of C11 with an area equal to that of C11. The inferred PD and EVPA for the C11 emission are consistent within the statistical uncertainties regardless of the background region used. We also investigated other procedures for the removal of intervals of high background. While the statistical significance of the polarization signal varies, the PD and EVPA after background subtraction remain consistent.

\begin{figure}
\hspace*{0.5cm}\includegraphics[width=0.42\textwidth]{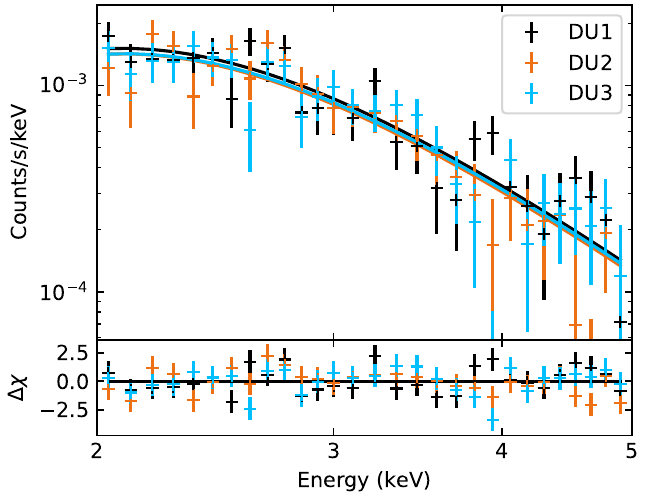}
\includegraphics[width=0.45\textwidth]{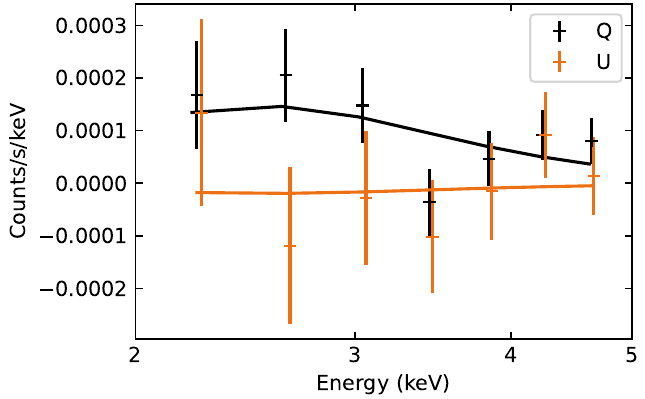}
\caption{IXPE X-ray spectra of the C11 region. (Top) Intensity (Stokes $I$) spectra. Data from the three IXPE DUs are shown as crosses: DU1 (black), DU2 (orange), and DU3 (blue). The fitted models, identical except for the normalization, are in the corresponding color. The lower panel shows the fit residuals.  (Bottom) Stokes $Q$ (black) and $U$ (orange) spectra. Points are shifted on the x-axis by $\pm 20$~eV for clarity. Data from the three DUs are added. The fitted model is shown as a solid line in the same color.} \label{fig:ixpe_spectrum}
\end{figure}

Figure~\ref{fig:ixpe_spectrum} shows the IXPE background-subtracted X-ray spectra for the C11 region. We performed spectropolarimetric analysis using Xspec. This differs from the \texttt{pcube} analysis in that it fully takes into account the response of the instrument including the non-diagonal portion of the energy response. The Stokes $I$ spectra have 80~eV bins below 3~keV and  120~eV above. The Stokes $Q$ and $U$ spectra have 400~eV bins. We modeled the spectrum using an absorbed power law with constant polarization multiplied by a constant allowed to vary between the three DUs. The column density for the \texttt{tbabs} absorption model using the \citet{Wilms2000} abundances was fixed to $1.18 \times 10^{22}\,\rm cm^{-2}$ as for the `Cone' region in \citet{SafiHarb2022}. The normalization for DU1 was fixed to unity and those for the other DUs were allowed to vary. We find a good spectropolarimetric fit ($\chi^2/\rm dof = 137.0/123$) with a photon index of $\Gamma = 1.57 \pm 0.14$, consistent with that found by \citet{SafiHarb2022}. The observed flux is $5.8 \times 10^{-13}\,\rm \, erg\,cm^{-2}\,s^{-1}$ in the 2--5~keV band. This corresponds to a luminosity of $1.3 \times 10^{34}\,\rm \,erg\,s^{-1}$ in the 0.3--30~keV band. The EVPA is $-3\fdg7 \pm  7\fdg8$, which is consistent with the \texttt{pcube} results. The PD is $52\% \pm 14\%$, which is consistent with the background-subtracted \texttt{pcube} result. The allowed range for PD considering the two analyses is 38\% to 77\%.


\section{Discussion}
\label{sec:discussion}

The high polarization degree found from the C11 region suggests the magnetic field has a well-ordered component, assuming that the X-rays are produced via synchrotron radiation. The maximum polarization for synchrotron radiation is $\Pi = (p+1)/(p+7/3)$ where the electron spectral index is $p = 2\Gamma - 1$. For the C11 region, $\Gamma = 1.57$ thus $p = 2.14$ and $\Pi = 0.70$. The ratio of the PD after background correction to the maximum polarization is approximately equal to the ratio of the energy in the magnetic field component with uniform direction to that in the total field  \citep{Burn1966}. For the C11 region, the lower bound on the PD of 38\% indicates at least $\approx 50$\% of the field is in the uniform component.

The energy of the electrons producing the emission can be calculated given the magnetic field strength. Assuming equipartition between the energies of the relativistic electrons and the magnetic field, the latter can be estimated from the X-ray luminosity and volume as $B \approx 10 \, \mu$G \citep{SafiHarb2022}. To produce X-rays of energy $E_{\gamma}$, the electron energy needed is $E_e \approx ({\rm 50 \, TeV}) (B/10 \, \mu {\rm G})^{-1/2} (E_\gamma/{\rm 1 \, keV})^{1/2}$. For $E_\gamma = 4 \,$keV and $B = 10\,\mu$G, electrons must be accelerated to 100~TeV. The same electrons also produce $\gamma$-rays near 25~TeV via inverse-Compton scattering on cosmic microwave background photons. Thus, IXPE probes the magnetic field configuration in the same region producing the VHE emission. The gyroradius for 100~TeV electrons in a $10\, \mu$G magnetic field is 0.011~pc which is much smaller than the transverse diameter of the C11 region of 3~pc. The synchrotron lifetime is 1200~yr \citep{Reynolds1998}. Thus, either the magnetic field is significantly higher than the equipartition estimate or some other loss mechanism dominates over the length of the region.

For synchrotron radiation, the EVPA is perpendicular to the magnetic field. Thus, the magnetic field lies along the direction of the main bulk motion of the jet. The key result of the IXPE observations is that the magnetic field near the acceleration region contains a significant component that is well-ordered and parallel to the flow direction. This result places constraints on models of the acceleration process and raises questions on the origin of the magnetic field.

The magnetic field could be intrinsic to the jet. However, while radio polarization measurements show the magnetic field is aligned with the ridge line of the precessing jet near SS~433 \citep{Stirling2004}, at larger distances, the field is instead parallel to the motion of individual knots \citep{MillerJones2008}. Thus, the field appears to be determined by the interaction between the individual knots and the ambient medium \citep{Blundell2018}. 

The powerful jets in Fanaroff-Riley Class II (FR II) radio galaxies have magnetic fields that are predominantly parallel to the jet axis all across the jet \citep{Bridle1984,Asseo1987}. The radio PDs are high, typically 20\%--40\% and reaching 50\%, e.g., in a knot in the jet of 4C 32.69 that may be analogous to the knot considered here \citep{Potash1980}. Velocity shear due to the interaction of the jet with the ambient medium is a natural means to produce a magnetic field aligned with the direction of bulk motion \citep{Blandford1983}.

Young, shell-type supernova remnants show radio EVPAs perpendicular to the flow near the shocks in the outer rims. Recent IXPE observations show similar EVPAs, typically with higher PDs, in the X-ray band \citep{Vink2022,Ferrazzoli2023,Zhou2023}. These results indicate magnetic fields parallel to the flow, as seen here. \citet{Giacalone2007} and \citet{Inoue2013} suggest that the magnetic field may stretch as the shock propagates around density fluctuations in the interstellar medium. This stretching can both amplify and align the field with the flow. Again, velocity shear is key to aligning the magnetic field with the flow. Alternatively, \citet{West2017} suggest that the overall field is disordered and that the mean EVPA represents the magnetic field only in localized regions where electrons are most efficiently accelerated because the magnetic field is parallel to the flow.

We consider several scenarios for acceleration and subsequent propagation and radiation of the energetic particles in the eastern lobe of SS~433. We note that the jet speed of 0.26$c$ implies a bulk Lorentz factor of 1.036. The magnetic field of $10 \, \mu$G estimated above and a jet density of $0.05 \rm \, cm^{-3}$ \citep{Panferov2017} implies an Alfv\'en Mach number of $\sim 900$ and a magnetization of $\sigma \sim 2 \times 10^{-4}$. Thus, the interaction is non-relativistic with a very high Mach number and the jet is hydrodynamical.

One possible scenario is stochastic acceleration in a shock \citep[see, e.g.,][]{Caprioli2014ApJ,Park2015PhRvL} at the leading edge of the e1 lobe followed by synchrotron emission from gradually cooling electrons flowing eastward \citep{Sudoh2020}. In this scenario, acceleration would be localized; \citet{SafiHarb2022} show a compact region at the leading edge that is bright in the 10-20~keV band and could be the acceleration region. Ordering of the parallel magnetic field would occur downstream of the turbulent shock region. The energetic electrons would suffer radiative loss while propagating downstream from the shock. Thus, the photon index of the nonthermal X-ray emission should gradually steepen eastward as is observed.

A second scenario is stochastic acceleration by turbulence \citep{Petrosian2012SSRv} in the shear interfaces between the jet and ambient medium. Such models have been proposed for FR~II radio galaxies and predict hard particle spectra consistent with those observed \citep{Stawarz2002,Rieger2004}. The acceleration is stochastic, hence the electron energy spectrum would be a power law with an index similar to that observed. Acceleration is not localized and can occur along the extent of the lobe where there is strong shear, thus steepening of the spectrum would not be expected over that region. The polarized emission would be predominantly from particles diffused out from the turbulent acceleration shear layer into the interior region with parallel magnetic field \citep{Tavecchio2021}.

A third scenario is acceleration through plasma processes \cite[see, e.g.,][]{Melrose1990AuJPh} associated with magnetic field reconnection. This may occur at magnetic loops produced by shear \citep{Romanova1992,Sironi2021} or at quasi-separatrix layers of adjacent magnetic flux ropes \citep[see][]{Aulanier2006SoPh}. The overall field reconfiguration process should be continual and gradual to maintain a quasi-equilibrium between particle acceleration, escape and radiative loss in order to produce a power-law spectrum with a relatively stable spectral index. While acceleration can occur wherever there is strong shear, the electron spectral index may vary due to local variations in the intensity of reconnection.

We note that Thomson scattering of radiation from the binary system, as alternative to synchrotron emission, would naturally produce highly polarized X-rays with an EVPA perpendicular to jet axis. Such a scenario requires an X-ray bright central source and SS~433 may be an ultraluminous X-ray source with the X-ray beam along the jet axis. \citet{Waisberg2019} place an upper bound on the X-ray luminosity of $\lesssim 5 \times 10^{36} \rm \, erg \, s^{-1}$ in a beam with a $1\arcdeg$ opening angle assuming the X-ray spectrum of a hard ultraluminous source \citep{Kaaret2017}. C11 viewed from SS~433 has an opening angle of $3\fdg2$, so the illuminating beam would have a luminosity of $\lesssim 5 \times 10^{37} \rm \, erg \, s^{-1}$. An optical depth of $3 \times 10^{-4}$ within C11 would be required to produce the observed luminosity, which is much larger than the expected optical depth of $3 \times 10^{-7}$ given the density estimate of \citet{Panferov2017}. We note that the differing morphology in different energy bands seen with NuSTAR \citep{SafiHarb2022} may also present issues with the scattering interpretation.

It may be possible to distinguish between these various models with future IXPE and other X-ray and multiwavelength observations. Deeper observations of the e1 lobe could enable spatially-resolved polarization measurements while observations of the e2 lobe could provide information on downstream evolution of the magnetic field. IXPE observation of the w1 lobe would be of great interest to determine if it also shows an EVPA perpendicular to the jet.

\begin{acknowledgments}
The Imaging X-ray Polarimetry Explorer (IXPE) is a joint US and Italian mission. The US contribution is supported by the National Aeronautics and Space Administration (NASA) and led and managed by its Marshall Space Flight Center (MSFC), with industry partner Ball Aerospace (contract NNM15AA18C). The Italian contribution is supported by the Italian Space Agency (Agenzia Spaziale Italiana, ASI) through contract ASI-OHBI-2017-12-I.0, agreements ASI-INAF-2017-12-H0 and ASI-INFN-2017.13-H0, and its Space Science Data Center (SSDC), and by the Istituto Nazionale di Astrofisica (INAF) and the Istituto Nazionale di Fisica Nucleare (INFN) in Italy.
A.V. acknowledges support from the Academy of Finland grant 355672.
\end{acknowledgments}

%
\facilities{IXPE, XMM-Newton, NuSTAR}

\software{
\textsc{ixpeobssim} \citep{baldnini22},
\textsc{xspec} \citep{xspec},
\textsc{ds9} \citep{ds9},
\textsc{astropy} \citep{astropy}
}

\bibliography{ss433refs}
\bibliographystyle{aasjournal}


\end{document}

%% file: author_list.tex
\author[0000-0002-3638-0637]{Philip Kaaret}
\affiliation{NASA Marshall Space Flight Center, Huntsville, AL 35812, USA}

\author[0000-0003-1074-8605]{Riccardo Ferrazzoli}
\affiliation{INAF Istituto di Astrofisica e Planetologia Spaziali, Via del Fosso del Cavaliere 100, 00133 Roma, Italy}

\author[0000-0002-8665-0105]{Stefano Silvestri}
\affiliation{Istituto Nazionale di Fisica Nucleare, Sezione di Pisa, Largo B. Pontecorvo 3, 56127 Pisa, Italy}

\author[0000-0002-6548-5622]{Michela Negro} 
\affiliation{Department of Physics and Astronomy, Louisiana State University, Baton Rouge, LA 70803, USA}

\author[0000-0002-0998-4953]{Alberto Manfreda}  
\affiliation{Istituto Nazionale di Fisica Nucleare, Sezione di Napoli, Strada Comunale Cinthia, 80126 Napoli, Italy}

\author[0000-0002-7568-8765]{Kinwah Wu}
\affiliation{Mullard Space Science Laboratory, University College London, Holmbury St Mary, Dorking, Surrey RH5 6NT, UK}

\author[0000-0003-4925-8523]{Enrico Costa}
\affiliation{INAF Istituto di Astrofisica e Planetologia Spaziali, Via del Fosso del Cavaliere 100, 00133 Roma, Italy}

\author[0000-0002-7781-4104]{Paolo Soffitta}
\affiliation{INAF Istituto di Astrofisica e Planetologia Spaziali, Via del Fosso del Cavaliere 100, 00133 Roma, Italy}

\author[0000-0001-6189-7665]{Samar Safi-Harb}
\affiliation{University of Manitoba, Department of Physics \& Astronomy, Winnipeg, MB R3T 2N2, Canada}

\author[0000-0002-0983-0049]{Juri Poutanen}
\affiliation{Department of Physics and Astronomy, 20014 University of Turku, Finland}

% ...
